# Evaluating $CO_2$ Storage Potential of Offshore Reservoirs and Saline Formations in Central Gulf of Mexico by Employing Data-driven Models with SAS® Viya

Rupom Bhattacherjee, Kodjo Opoku Botchway, Xitong Hu, Jack Pashin, Goutam Chakraborty, and Prem Bikkina, Oklahoma State University

## ABSTRACT

The SECARB offshore partnership project seeks to screen deep saline aquifers and hydrocarbon reservoirs in the central Gulf of Mexico (GOM) for $CO_2$ sequestration and $CO_2$-driven enhanced oil and gas recovery and estimate the corresponding $CO_2$ storage resources for select reservoirs. To this end, three major objectives have been completed: managing geological data from different sources, building a reservoir screening platform for $CO_2$ storage, and ranking the reservoirs based on the estimated storage potential. First, the major geological characteristics of both shelf and deep-water areas of the central GOM were examined and compared to define the appropriate reservoir screening criteria. Consequently, the $CO_2$ storage resources of the screened reservoirs were calculated and reported at the BOEM field level to identify fields with the highest storage potential. In the project's current phase, the assessment is being expanded to saline formations. Correlations are being identified, tested, and developed for a broad range of rock and fluid properties, including thickness, porosity, permeability, fluid saturation, and fluid chemistry. These correlations are developed for interfacial tension and $CO_2$ saturated brine viscosity to improve the storage estimates and consider the capillary trapping and solubility of $CO_2$ in reservoir fluids. SAS® Viya software is used to make visualizations of these properties of the offshore reservoir and make comparisons and draw similarities between the experimental and estimated correlated measures. Using interactive plots, different conditions could be screened, making the analysis more attractive from a user's point of view.

## INTRODUCTION

The substantial $CO_2$ storage potential of the hydrocarbon reservoirs and saline formations in the central Gulf of Mexico (continental shelf and continental slope offshore of Louisiana) makes future development of $CO_2$ storage projects in those areas promising for a large-scale, long-term, and secure storage of $CO_2$. However, subsurface storage of $CO_2$ is always more complicated than merely locking the gas in a sealed container, hence requires assessing the geological properties of the underground formations and the thermophysical properties of the formation fluids (e.g., $CO_2$, brine).

The formation fluids and the underground formations need to qualify in several aspects to ensure safe and long-term storage of $CO_2$ (Pijaudier-Cabot and Pereira, 2013). For example, the formation fluid needs to have high $CO_2$ solubility to favor the solubility trapping, one of the major mechanisms of $CO_2$ getting trapped in the subsurface formation; the sealing capillary pressure, which is a function of $CO_2$-brine interfacial tension, needs to be higher than the buoyancy forces exerted by the fluids underlying the caprock to prevent the upward migration and escape of stored $CO_2$; the viscosity contrast between $CO_2$ and $CO_2$ saturated brine needs to be sufficiently high to prevent the displacement of stored $CO_2$ by brine; the storage unit should be sufficiently porous and permeable to hold the gas; the reservoirs should have enough headroom to inject the $CO_2$. Designing and optimizing subsurface $CO_2$ storage thus requires a good understanding of the thermo-physical properties of the $CO_2$ and formation brine as well as the geological properties of the target formation.

This study is part of the SECARB Offshore Partnership Project (DE-FE0031557) to evaluate the $CO_2$ storage potentials of the hydrocarbon reservoirs and saline formations in the central Gulf of Mexico (GOM). The study estimates a few significantly important thermophysical properties of $CO_2$ and brine for the central GOM sands. These include the interfacial tension between $CO_2$ and brine, the solubility of $CO_2$ in brine, and the viscosity of $CO_2$ and $CO_2$ saturated brine. In addition to understanding the fluid behaviors in those formations, this study will help estimate the trapping and solubility potential of the reservoirs.

## STUDY AREA AND DATA OVERVIEW

The study uses the Bureau of Ocean Energy Management's (BOEM) 2019 Sands Atlas dataset (www.data.boem.gov). The dataset includes detailed information on the location, reservoir properties, well log measurements, core measurements, and production history at individual wells and their productive sand levels. The data here was segregated into the shelf and slope areas based on the protractions of the individual fields, but for this study, the information present is from the slope area only. The minimum, maximum, and average initial reservoir temperatures of the studied reservoirs were 21.1 °C, 164.4 °C, and 76.8 °C, respectively. The pressures of the fields have a minimum initial reservoir pressure of 26.96 bar, a maximum of 1489.88 bar, and an average of 385.13 bar.

## METHODOLOGY AND MODELLING

### INTERFACIAL TENSION

Interfacial tension (IFT) is the force of attraction between the molecules at the interface of two fluids. IFT determines the separation potential to mix between two fluids or phases in contact, which in the case of subsurface $CO_2$ storage would be the brine/oil and the $CO_2$ inside the rock (Moghadasi, Rostami et al. 2018). This intrinsic property of the interaction primarily affects the amounts of $CO_2$ that can be stored in the reservoir due to structural trapping. The IFT between the brine and the $CO_2$ can be affected by several parameters, including the cation valance molality of the brine, reservoir temperature, pressure, and the concentrations of the brine.

This study used the correlation developed by Jerauld & Kazemi (Jerauld and Kazemi 2022) to estimate the $CO_2$-brine IFT for the BOEM reservoirs mentioned in the introduction section. The execution of this correlation was expressed as a function of the density difference between water-saturated $CO_2$ and $CO_2$-saturated brine, temperature, and valence-weighted molality of cations (e.g., Na+).

$CO_2$ density was estimated as a function of temperature and pressure using the Spycher et al. $CO_2$ density prediction method (Spycher, Pruess et al. 2003), modified by Bikkina et al. to evaluate the saturated $CO_2$ phase densities (Bikkina, Shoham et al. 2011). The density of $CO_2$-saturated brine was estimated as a function of temperature, pressure, and mass fraction of $CO_2$ in brine using the model developed by Bachu and Adams (Bachu and Adams 2003), with the corresponding $CO_2$ solubility data estimated using the solubility model developed by Duan and Sun (Duan and Sun 2003).

A total of 1717 experimental data on $CO_2$-brine IFT was collected from previous studies (Ren, Chen et al. 2000, Yan, Zhao et al. 2001, Hebach, Oberhof et al. 2002, Chiquet, Daridon et al. 2007, Kvamme, Kuznetsova et al. 2007, Bachu and Bennion 2009, Chalbaud, Robin et al. 2009, Aggelopoulos, Robin et al. 2010, Georgiadis, Maitland et al. 2010, Aggelopoulos, Robin et al. 2011, Bikkina, Shoham et al. 2011, Li, Boek et al. 2012, Li, Boek et al. 2012) that include the IFT between pure/impure $CO_2$ and pure water/brine of different composition and salinities, and were used to validate the estimations from the model used in this study. The set of equations used to estimate IFT between $CO_2$ and brine are as follows:

**Interfacial Tension Estimation (Jerauld and Kazemi, 2022)**

$$\Upsilon = cn + df\sqrt{\Delta\rho} + pf(\Delta\rho)^{ex} + a[m^+] \qquad (1)$$
$$cn = cn_0 + cn_1 T_r \qquad (2)$$
$$df = df_0 + df_1 T_r + df_2 T_r^2 \qquad (3)$$
$$ex = ex_0 + ex_1 T_r + ex_2 T_r^2 \qquad (4)$$
$$pf = pf_0 + pf_1 T_r + pf_2 T_r^2 \qquad (5)$$
$$a = a_0 + a_1 T_r + a_2 T_r^{-1} \qquad (6)$$

Where, ϒ stands for the IFT between the $CO_2$ and brine, mN/m; Δρ is the density difference, kg/m³; $T_r$ is the reduced temperature; m+ is the valence-weighted molality of the cations, mol/kg.

**Table 1. Parameters and their values for Eqs. 2-6**

| Parameter | Values | Parameter | Values | Parameter | Values | Parameter | Values |
|---|---|---|---|---|---|---|---|
| $cn_1$ | -41.146 | $df_0$ | -147.856 | $pf_3$ | 259.490 | $a_2$ | -0.9311 |
| $cn_0$ | 68.790 | $ex_2$ | 15.071 | $pf_2$ | -930.066 | $a_1$ | 0.4753 |
| $df_2$ | -91.281 | $ex_1$ | -52.516 | $pf_1$ | 1083.514 | $a_0$ | 2.0181 |
| $df_1$ | 245.472 | $ex_0$ | 49.239 | $pf_0$ | -374.117 | | |

## VISCOSITY OF $CO_2$ AND $CO_2$-SATURATED BRINE

For any system incorporating a multiphase flow system, there exists the importance of modeling the viscosity of the fluid phases (Peter, Yang et al. 2022). The viscosity of $CO_2$ and bine affect the mobility of the fluids and may cause issues such as viscous fingering and low volumetric efficiency if the injected $CO_2$ has low viscosity (Yu, Lashgari et al. 2015). The viscosity of the injected $CO_2$ phase must be adequate to maintain the balance that seeks to keep the $CO_2$/$CO_2$-brine phase immovable. $CO_2$ viscosity also determines the energy and economic feasibility of $CO_2$ transportation through pipelines, one of the major ways of transporting $CO_2$ to storage sites in the USA and Canada (Cole and Itani 2013).

Brine viscosity is directly related to its density which may vary depending on the dissolution of $CO_2$ (Duan and Sun 2003). In the case of storing $CO_2$ underground, dissolution of $CO_2$ in brine is expected; hence the viscosity estimation of brine should take $CO_2$ solubility into account. In this study, we estimated the viscosity of $CO_2$ and $CO_2$ saturated NaCl brine using a combination of equations of states (EOS). $CO_2$ viscosity was estimated using a recent empirical model developed by Nait Amar et al. (Amar, Ghriga et al. 2020). The model is applicable in the temperature and pressure ranges from 220–673 K and 0.1 MPa to 7960 MPa.

Estimating $CO_2$ saturated NaCl brine required estimating pure water density, the viscosity of brine (NaCl+$H_2O$), and the mole fraction of $CO_2$ in brine. The density of the water is obtained from the formulation developed by the International Association for the Properties of Water and Steam (Wagner and Kretzschmar 2008). The viscosity of the $H_2O$+NaCl system was estimated using the dynamic viscosity model developed by Mao and Duan (Mao and Duan 2009). The $CO_2$ mole fraction is calculated using the Duan and Sun model for $CO_2$ solubility in brine (Duan and Sun 2003). The model developed was valid up to 273 to 533 K, 0-200 MPa, and 0–6.5 m NaCl. Finally, the viscosity of $CO_2$ saturated NaCl brine is estimated with the empirical correlation developed by Islam and Carlson (Islam and Carlson 2012).

The models were selected based on their performance in accurately estimating the experimental values of $CO_2$ viscosity (Li, Wilhelmsen et al. 2011, Laesecke and Muzny 2017), NaCl+$H_2O$ (Kestin, Khalifa et al. 1977, Ozbek, Fair et al. 1977, Semenyuk, Zarembo et al. 1977, Kestin, Khalifa et al. 1978, Kestin, Khalifa et al. 1981, Kestin and Shankland 1984), $CO_2$ mole fraction (Rumpf, Nicolaisen et al. 1994, Bando, Takemura et al. 2004, Messabeb, Contamine et al. 2016, Mohammadian, Liu et al. 2022), and $CO_2$-saturated brine viscosity (Bando, Takemura et al. 2004) over the temperature and pressure range of the BOEM fields.

The summary of the equations used in the model development is shown below:

### $CO_2$ Viscosity (Amar, Ghriga et al. 2020)

$$\mu_{CO_2} = [0.5703 + 0.01033\sqrt{T-11.69} - 0.0001304 \times A + 0.134 \times B + 10^{-5} \times \sqrt{\rho} \times C]^{10} \quad (7)$$

In this equation, T and ρ are temperature and $CO_2$ density, respectively. $CO_2$ density is estimated using NIST Chemistry WebBook (Laesecke and Muzny 2017). The constants A, B, and C are defined as follows:

$$A = \rho + T + \ln(\rho) \quad (8)$$

$$B = \tanh\left(\frac{T}{\rho}\right) - 0.271194 \times \tanh\left(\frac{\rho}{T}\right) - 0.063664 \times \tanh(T-\rho) \quad (9)$$

$$C = 1.624 \times (T + \rho) - 2.363 \times \sqrt{T \times \rho} - 480.8 \tag{10}$$

T and ρ should be expressed in units of K and kg/m3, respectively, to obtain viscosity, $\mu_{CO_2}$ in mPa.s.

## NaCl+H$_2$O Viscosity (Mao and Duan 2009)

$$\mu_{brine} = \mu_{relative} \times \mu_{pw} \tag{11}$$

$$\mu_{relative} = e^{(A \times m + B \times m^2 + C \times m^3)} \tag{12}$$

$$\log_{10} \mu_{pw} = d_i[1] \times (T + 273.15)^{1-3} + d_i[2] \times (T + 273.15)^{2-3} + d_i[3] \times (T + 273.15)^{3-3} + d_i[4] \times (T + 273.15)^{4-3} + d_i[5] \times (T + 273.15)^{5-3} + d_i[6] \times \left(\frac{\rho_{pw}}{1000}\right) \times (T + 273.15)^{6-8} + d_i[7] \times \left(\frac{\rho_{pw}}{1000}\right) \times (T + 273.15)^{7-8} + d_i[8] \times \left(\frac{\rho_{pw}}{1000}\right) \times (T + 273.15)^{8-8}) + d_i[9] \times \left(\frac{\rho_{pw}}{1000}\right) \times (T + 273.15)^{9-8} + d_i[6] \times \left(\frac{\rho_{pw}}{1000}\right) \times (T + 273.15)^{6-8} \tag{13}$$

In equation 11, $\mu_{relative}$ is the relative viscosity of brine to pure water; m in equation 12 is the molality of brine. Other constants can be calculated as follows:

$$A = a[0] + a[1] \times (T + 273.15) + a[2] \times (T + 273.15)^2 \tag{14}$$

$$B = b[0] + b[1] \times (T + 273.15) + b[2] \times (T + 273.15)^2 \tag{15}$$

$$C = c[0] + c[1] \times (T + 273.15) \tag{16}$$

The values for constants a, b, c, and di in equations 13 to 16 are listed in table 1. The temperature should be expressed in °C, pressure in bar, density in g/cc, molality in mol/kg, and viscosity in mPa.s.

| Par. | Values | Par. | Values | Par. | Values | Par. | Values | Par. | Values |
|---|---|---|---|---|---|---|---|---|---|
| a$_0$ | -0.21 | b$_1$ | 0.27×10$^{-3}$ | d$_{i0}$ | 0 | d$_{i4}$ | -0.31×10$^{-1}$ | d$_{i7}$ | 0.56 ×10$^4$ |
| a$_1$ | 0.17×10$^{-2}$ | b$_2$ | 0.21×10$^{-6}$ | d$_{i1}$ | 0.28×10$^7$ | d$_{i5}$ | -0.27×10$^{-4}$ | d$_{i8}$ | 0.14×10$^8$ |
| a$_2$ | 0.12×10$^{-5}$ | c$_0$ | -0.26×10$^{-2}$ | d$_{i2}$ | -0.11×10$^5$ | d$_{i6}$ | -0.19×10$^7$ | d$_{i9}$ | 0.48×10$^1$ |
| b$_0$ | 0.69×10$^{-1}$ | c$_1$ | 0.78×10$^{-5}$ | d$_{i3}$ | -0.91×10$^1$ | d$_{i6}$ | -0.19×10$^7$ | d$_{i10}$ | 0.35×10$^{-4}$ |

**Table 2. Parameters and their values for Eqs. 14-16**

## Viscosity of H$_2$O+NaCl+CO$_2$ (Islam and Carlson 2012)

$$\mu_{H_2O+NaCl+CO_2} = \mu_{brine}(1 + 4.65 \times m_{CO_2}^{1.0134}) \tag{17}$$

$\mu_{brine}$, in equation (17), is the viscosity of the H$_2$O+NaCl system, estimated using equation (11). m$_{CO2}$ is the mole fraction of CO$_2$ in brine, calculated using the solubility model developed by Duan and Sun (Duan and Sun 2003).

## CO$_2$ SOLUBILITY IN NACL BRINE

One of the major considerations for successful long-term safety of CO$_2$ storage is to anticipate the spread of CO$_2$ in the aquifer (Riaz and Cinar 2014). For that to be appropriately ascertained, the solubility of the CO$_2$ in the formation water (brine) needs to be known. This is essential to identify the potential for free-phase CO$_2$ or brine leakage through the fault fractures in the subsurface structures (Meguerdijian, Pawar et al. 2022). To arrive at that, the balance between the potential of the CO$_2$ in the liquid phase and that in the gas phase would be needed to determine the CO$_2$ solubility in aqueous solutions. This is represented as the calculation of fugacity ($\phi_{CO_2}$). From this approximation, the solubility ($m_{CO_2}$) can be estimated as follows:

$$ln \frac{\Upsilon_{CO_2} P}{m_{CO_2}} = \frac{\mu_{CO_2}^{1(0)}}{RT} - ln\phi_{CO_2} + \sum_c 2\lambda_{CO_2-c} m_c + \sum_a 2\lambda_{CO_2-a} m_a + \sum_c \sum_a \zeta_{CO_2-a-c} m_c m_a \qquad (18)$$

In the equation (18), the λ and the ζ are the second and third order interaction parameters. The $\frac{\mu_{CO_2}^{1(0)}}{RT}$ term is the dimensionless standard chemical potential (fugacity potential) and are dependent on the temperature and pressure of the system.

## RESULTS AND DISCUSSION

### MODEL PERFORMANCE

The section is a summary of the performance of the models with respect to the data that was fed into it. The model predictions and the experimental results are being compared here, and the results are shown below.

### $CO_2$-Brine Interfacial Tension

Fig 1a shows the comparison chart between the experimental and estimated IFTs. The selected model for IFT performed well on the experimental data. For the same density difference, experimental and estimated IFT values fell on top of each other, especially at higher density differences (>0.5 g/cc). There are some instances in the low-density difference region where the estimated values are not matching very well with the experimental values. That could be due to the phase change of the $CO_2$ when the temperature and pressure are above the critical region (Bikkina, Shoham et al. 2011). Nevertheless, the model estimated the IFTs with great accuracy as the mean absolute error (MAE) was only 1.29 mN/m.

### $CO_2$ Viscosity

$CO_2$ viscosity was estimated as a function of T, P, and $CO_2$ density. The experimental $CO_2$ viscosity values are plotted against the estimated ones. The 45° reference line is drawn to show the deviation of estimated values from the experimental data (Fig. 1b). The estimated values fell very close to the experimental ones. The MAE was only 0.23 mPa.s, indicating an excellent accuracy of the model in estimating the viscosity of $CO_2$.

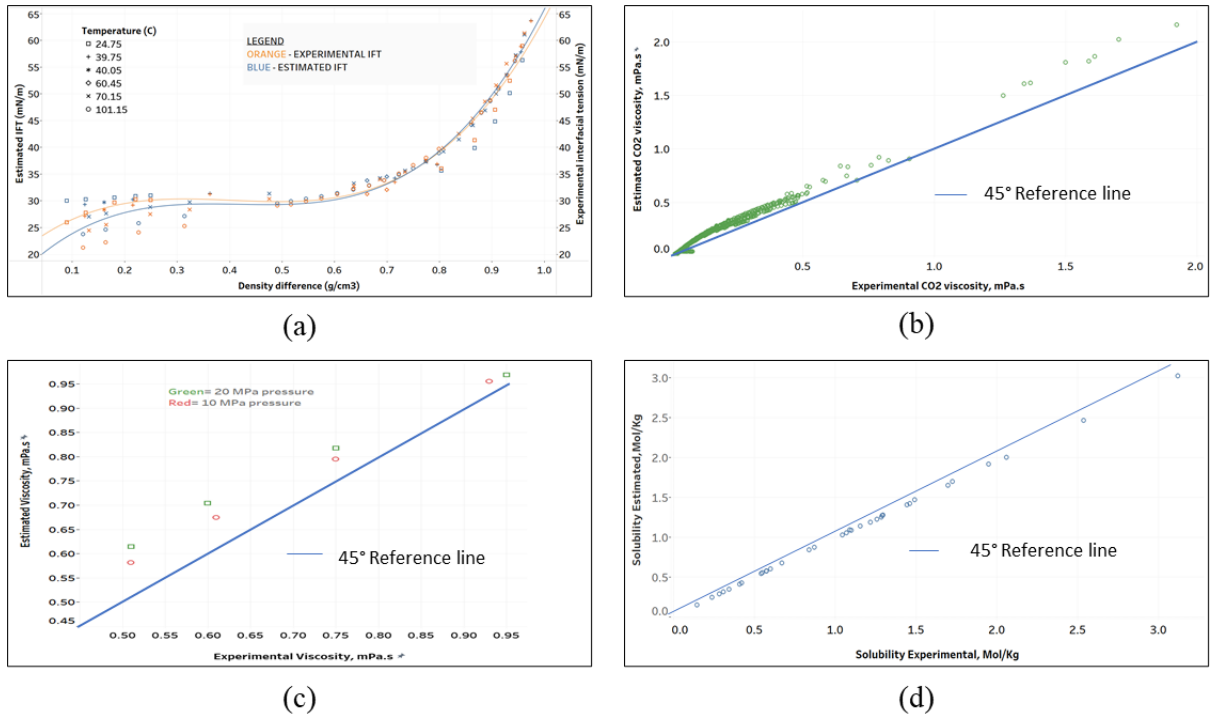

**Figure 1:** Comparison between the experimental and estimated values of (a) IFT between $CO_2$ and NaCl brine, viscosities of (b) $CO_2$ and (c) $CO_2$ saturated NaCl brine and (d) solubility of $CO_2$ in NaCl brine

### Viscosity of $H_2O$+NaCl+$CO_2$

Unlike the $CO_2$ viscosity, in the case of $CO_2$ saturated brine, there were not enough experimental data to validate the model. The only relevant data source we found covering the range of temperature and pressure usually encountered in the geological formations for $CO_2$ storage is from Bando et al. (Bando, Takemura et al. 2004). The comparison of the estimated viscosity with their data is shown in figure 1c. Even though it appears from the plot that the model is significantly overestimating the viscosities, the estimates are only slightly different from the experimental values, as the MAE was found to be only 0.06 mPa.s. Due to the representation of the units being in the tenths, a difference of 0.06 units of measurement shows a significant visual difference in the plot.

### $CO_2$ Solubility in Brine

Fig 1d shows the comparison between the model estimated solubility compared with the experimental values obtained. The model estimated the $CO_2$ solubility as a function of T, P, brine composition (cation valence molality), and water-saturated $CO_2$ density. The estimated mean absolute error of this model was 0.01 mol/kg.

### ESTIMATIONS FOR THE STUDY AREA

The section is a representation of the $CO_2$-brine IFT, viscosity of $CO_2$ and $CO_2$ saturated brine, and the solubility of $CO_2$ in brine estimated for the reservoirs from the central Gulf of Mexico (BOEM sands) using the models mentioned in section 3.

Figure 2 shows the spread of IFT values throughout the BOEM sands on the slope region. Red indicates less IFT, while blue indicates very high $CO_2$ brine interfacial tension. As can be seen, the IFT ranges from 14.15 to 68.08 mN/m, and very few fields have a very high $CO_2$-brine IFT. Figure 3 and figure 4 show the field spread of the viscosity of $CO_2$ and $CO_2$ saturated brine, respectively. While the $CO_2$ viscosity in the central GOM reservoirs ranges from 0.02 to 0.16 mPa.s, the brine viscosity ranges from 0.16 to 1.34 mPa.s.

However, in both cases, the Northern boundary sands (marked blue) are attributed to lower viscosities than the southern ones (marked red).

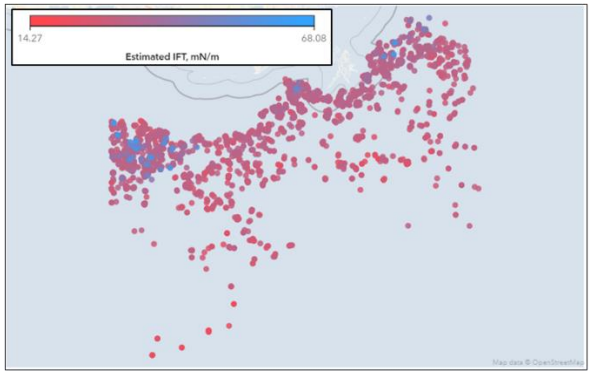 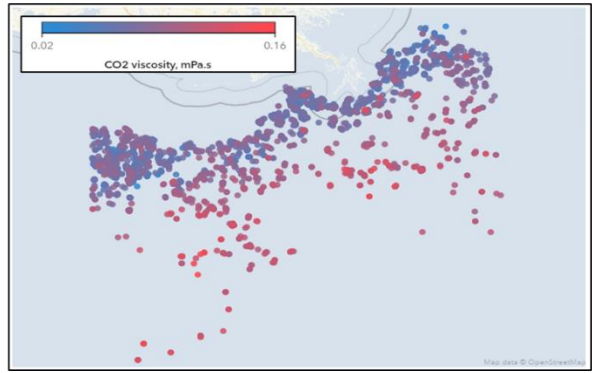

**Fig. 2: Field spread of the estimated IFT between $CO_2$ and NaCl brine for the study area**

**Fig. 3: Field spread of the estimated $CO_2$ Viscosity**

The $CO_2$ solubility potential of the reservoirs is shown in figure 5. The solubility ranges from 0.6 to 1.4 mol/kg in brine. Sands in the central, west and south-west regions (marked deep green) are estimated to have very high solubility potential, close to 1 mol/kg. Table 2 summarizes the findings of all the thermo-physical properties estimated in this study for the BOEM sands. It also includes the estimated values of the properties such as the density of $H_2O$, $NaCl+H_2O$, $CO_2$, and $NaCl+H_2O+CO_2$ that were used to estimate the other properties.

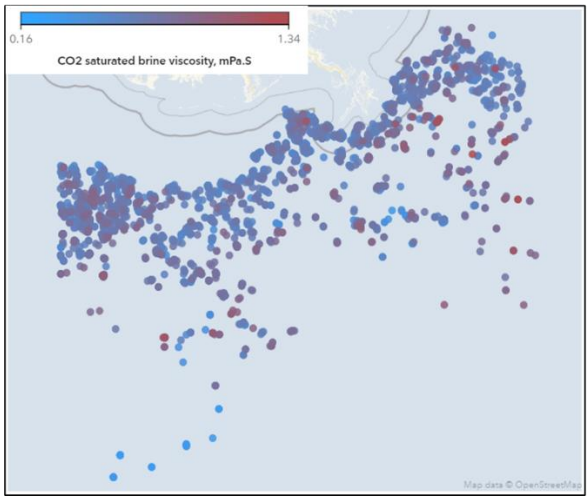 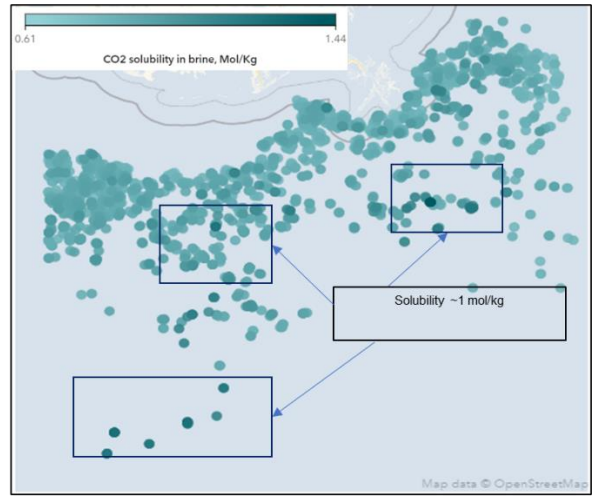

**Fig. 4: Field spread of the estimated viscosity of $H_2O+NaCl+CO_2$ for the study area in the central GOM**

**Fig. 5: Field spread of the estimated $CO_2$ solubility in NaCl brine for the study area in the central GOM**

**Table 3: Summary of the properties estimated for the study area in the central GOM**

|  | Interfacial Tension (mN/m) | $CO_2$ Saturated Brine Viscosity (mPa.s) | $CO_2$ Viscosity (mPa.s) | $CO_2$ solubility (mol/kg) | Water Density (kg/m3) | $CO_2$ Brine Density (kg/m3) | $CO_2$ Density (kg/m3) |
|---|---|---|---|---|---|---|---|
| Min | 14.145 | 0.1595 | 0.0162 | 0.6133 | 55.22 | 1.0301 | 0.0552 |
| Max | 68.081 | 1.3446 | 0.1642 | 1.4392 | 1105.98 | 1.2559 | 1.1059 |
| Avg | 29.984 | 0.6442 | 0.0731 | 0.8463 | 756.52 | 1.1090 | 0.7565 |

## CONCLUSION

This paper was made to report on the importance of these thermo-physical properties of the $CO_2$ and the brine phases, along with their interactions with respect to geologic storage. The consolidated models were validated against the experimental values to be able to make profiles of the properties. These consolidated models built can make reasonable estimations of the IFT, $CO_2$ viscosity, $CO_2$ saturated brine viscosity, and $CO_2$ solubility in brine. IFT and $CO_2$ solubility data can be used to estimate the trapping and solubility potential of the reservoirs for $CO_2$ storage, respectively, while the viscosity data can help understand the displacement efficiency of $CO_2$ and brine in the reservoirs as well as the transport properties of $CO_2$ in the pipelines.

## REFERENCES

Aggelopoulos, C., M. Robin and O. Vizika (2011). "Interfacial tension between CO2 and brine (NaCl+ CaCl2) at elevated pressures and temperatures: The additive effect of different salts." Advances in Water Resources 34(4): 505-511.

Aggelopoulos, C. A., M. Robin, E. Perfetti and O. Vizika (2010). "CO2/CaCl2 solution interfacial tensions under CO2 geological storage conditions: influence of cation valence on interfacial tension." Advances in Water Resources 33(6): 691-697.

Amar, M. N., M. A. Ghriga, H. Ouaer, M. E. A. B. Seghier, B. T. Pham and P. Ø. Andersen (2020). "Modeling viscosity of CO2 at high temperature and pressure conditions." Journal of Natural Gas Science and Engineering 77: 103271.

Bachu, S. and J. J. Adams (2003). "Sequestration of CO2 in geological media in response to climate change: capacity of deep saline aquifers to sequester CO2 in solution." Energy Conversion and management 44(20): 3151-3175.

Bachu, S. and D. B. Bennion (2009). "Interfacial tension between CO2, freshwater, and brine in the range of pressure from (2 to 27) MPa, temperature from (20 to 125) C, and water salinity from (0 to 334 000) mg· L− 1." Journal of Chemical & Engineering Data 54(3): 765-775.

Bando, S., F. Takemura, M. Nishio, E. Hihara and M. Akai (2004). "Viscosity of aqueous NaCl solutions with dissolved CO2 at (30 to 60) C and (10 to 20) MPa." Journal of Chemical & Engineering Data 49(5): 1328-1332.

Bikkina, P. K., O. Shoham and R. Uppaluri (2011). "Equilibrated interfacial tension data of the CO2–water system at high pressures and moderate temperatures." Journal of Chemical & Engineering Data 56(10): 3725-3733.

Chalbaud, C., M. Robin, J. Lombard, F. Martin, P. Egermann and H. Bertin (2009). "Interfacial tension measurements and wettability evaluation for geological CO2 storage." Advances in water resources 32(1): 98-109.

Chiquet, P., J.-L. Daridon, D. Broseta and S. Thibeau (2007). "CO2/water interfacial tensions under pressure and temperature conditions of CO2 geological storage." Energy Conversion and Management 48(3): 736-744.

Cole, S. and S. Itani (2013). "The Alberta carbon trunk line and the benefits of CO2." Energy Procedia 37: 6133-6139.

Duan, Z. and R. Sun (2003). "An improved model calculating CO2 solubility in pure water and aqueous NaCl solutions from 273 to 533 K and from 0 to 2000 bar." Chemical geology 193(3-4): 257-271.

Georgiadis, A., G. Maitland, J. M. Trusler and A. Bismarck (2010). "Interfacial tension measurements of the (H2O+ CO2) system at elevated pressures and temperatures." Journal of Chemical & Engineering Data 55(10): 4168-4175.

Hebach, A., A. Oberhof, N. Dahmen, A. Kögel, H. Ederer and E. Dinjus (2002). "Interfacial tension at elevated pressures measurements and correlations in the water+ carbon dioxide system." Journal of Chemical & Engineering Data 47(6): 1540-1546.

Islam, A. W. and E. S. Carlson (2012). "Viscosity models and effects of dissolved CO2." Energy and Fuels: 5330 - 5336.

Islam, A. W. and E. S. Carlson (2012). "Viscosity models and effects of dissolved CO2." Energy & fuels 26(8): 5330-5336.

Jerauld, G. R. and A. Kazemi (2022). "An improved simple correlation for accurate estimation of CO2-Brine interfacial tension at reservoir conditions." Journal of Petroleum Science and Engineering: 208.

Kestin, J., H. E. Khalifa, Y. Abe, C. E. Grimes, H. Sookiazian and W. A. Wakeham (1978). "Effect of pressure on the viscosity of aqueous sodium chloride solutions in the temperature range 20-150. degree. C." Journal of Chemical and Engineering Data 23(4): 328-336.

Kestin, J., H. E. Khalifa and R. J. Correia (1981). "Tables of the dynamic and kinematic viscosity of aqueous NaCl solutions in the temperature range 20–150 C and the pressure range 0.1–35 MPa." Journal of physical and chemical reference data 10(1): 71-88.

Kestin, J., H. E. Khalifa, S.-T. Ro and W. A. Wakeham (1977). "Preliminary data on the pressure effect on the viscosity of sodium chloride-water solutions in the range 10-40. degree. C." Journal of Chemical and Engineering Data 22(2): 207-214.

Kestin, J. and I. R. Shankland (1984). "Viscosity of aqueous NaCl solutions in the temperature range 25–200 C and in the pressure range 0.1–30 MPa." International Journal of Thermophysics 5(3): 241-263.

Kvamme, B., T. Kuznetsova, A. Hebach, A. Oberhof and E. Lunde (2007). "Measurements and modelling of interfacial tension for water+ carbon dioxide systems at elevated pressures." Computational Materials Science 38(3): 506-513.

Laesecke, A. and C. D. Muzny (2017). "Reference correlation for the viscosity of carbon dioxide." Journal of physical and chemical reference data 46(1): 013107.

Li, H., Ø. Wilhelmsen, Y. Lv, W. Wang and J. Yan (2011). "Viscosities, thermal conductivities and diffusion coefficients of CO2 mixtures: Review of experimental data and theoretical models." International Journal of Greenhouse Gas Control 5(5): 1119-1139.

Li, X., E. Boek, G. C. Maitland and J. M. Trusler (2012). "Interfacial Tension of (Brines+ CO2):(0.864 NaCl+ 0.136 KCl) at Temperatures between (298 and 448) K, Pressures between (2 and 50) MPa, and Total Molalities of (1 to 5) mol· kg–1." Journal of Chemical & Engineering Data 57(4): 1078-1088.

Li, X., E. S. Boek, G. C. Maitland and J. M. Trusler (2012). "Interfacial Tension of (Brines+ CO2): CaCl2 (aq), MgCl2 (aq), and Na2SO4 (aq) at Temperatures between (343 and 423) K, Pressures between (2 and 50) MPa, and Molalities of (0.5 to 5) mol· kg–1." Journal of Chemical & Engineering Data 57(5): 1369-1375.

Mao, S. and Z. Duan (2009). "The viscosity of aqueous alkali-chloride solutions up to 623 K, 1,000 bar, and high ionic strength." International Journal of Thermophysics: 1510 - 1523.

Mao, S. and Z. Duan (2009). "The viscosity of aqueous alkali-chloride solutions up to 623 K, 1,000 bar, and high ionic strength." International Journal of Thermophysics 30(5): 1510-1523.

Meguerdijian, S., R. J. Pawar, D. R. Harp and B. Jha (2022). "Thermal and solubility effects on fault leakage during geologic carbon storage." International Journal of Greenhouse Gas Control.

Messabeb, H., F. Contamine, P. Cézac, J. P. Serin and E. C. Gaucher (2016). "Experimental Measurement of CO2 Solubility in Aqueous NaCl Solution at Temperature from 323.15 to 423.15 K and Pressure of up to 20 MPa." Journal of Chemical & Engineering Data 61(10): 3573-3584.

Moghadasi, R., A. Rostami and A. Hemmati-Sarapardeh (2018). Chapter Three - Enhanced Oil Recovery Using CO2. Fundamentals of Enhanced Oil and Gas Recovery from Conventional and Unconventional Reservoirs. Lismore, NSW, Australia: 61 - 99.

Mohammadian, E., B. Liu and A. Riazi (2022). "Evaluation of Different Machine Learning Frameworks to Estimate CO2 Solubility in NaCl Brines: Implications for CO2 Injection into Low-Salinity Formations." Lithosphere 2022(Special 12): 1615832.

Ozbek, H., J. Fair and S. Phillips (1977). Viscosity of Aqueous Sodium Chloride Solutions From 0-150o c, Lawrence Berkeley National Lab.(LBNL), Berkeley, CA (United States).

Peter, A., D. Yang, K. I.-I. I. Eshiet and Y. Sheng (2022). "A Review of the Studies on CO2–Brine–Rock Interaction in Geological Storage Process." Geosciences.

Ren, Q.-Y., G.-J. Chen, W. Yan and T.-M. Guo (2000). "Interfacial tension of (CO2+ CH4)+ water from 298 K to 373 K and pressures up to 30 MPa." Journal of Chemical & Engineering Data 45(4): 610-612.

Riaz, A. and Y. Cinar (2014). "Carbon dioxide sequestration in saline formations: Part I—Review of the modeling of solubility trapping." Journal of Petroleum Science and Engineering: 367 - 380.

Rumpf, B., H. Nicolaisen, C. Öcal and G. Maurer (1994). "Solubility of carbon dioxide in aqueous solutions of sodium chloride: experimental results and correlation." Journal of solution chemistry 23(3): 431-448.

Semenyuk, E. N., V. I. Zarembo and M. K. Fedorov (1977). "Apparatus for measuring viscoisties of electrolyte-solutions at temperatures of 273-673 K and pressures up to 200 MPa." J. Appl. Chem. USSR 50: 298-302.

Spycher, N., K. Pruess and J. Ennis-King (2003). "CO2-H2O mixtures in the geological sequestration of CO2. I. Assessment and calculation of mutual solubilities from 12 to 100 C and up to 600 bar." Geochimica et cosmochimica acta 67(16): 3015-3031.

Wagner, W. and H.-J. Kretzschmar (2008). "IAPWS industrial formulation 1997 for the thermodynamic properties of water and steam." International steam tables: properties of water and steam based on the industrial formulation IAPWS-IF97: 7-150.

Yan, W., G.-Y. Zhao, G.-J. Chen and T.-M. Guo (2001). "Interfacial tension of (methane+ nitrogen)+ water and (carbon dioxide+ nitrogen)+ water systems." Journal of Chemical & Engineering Data 46(6): 1544-1548.

Yu, W., H. R. Lashgari, K. Wu and K. Sepehrnoori (2015). "CO2 injection for enhanced oil recovery in Bakken tight oil reservoirs." Fuel 159: 354-363.

# ACKNOWLEDGMENTS

The National Energy Technology Laboratory (NETL) of the United States Department of Energy (DOE) provided funding (Award No. DE-FE0031557) for this study through the Southern States Energy Board's Southeast Regional Carbon Storage Partnership (SECARB Offshore) project. Cost share supporting this research was supplied by the project partners, including the SAS Institute. This paper is based upon work supported by the Department of Energy and was prepared as an account of work sponsored by an agency of the United States Government. Neither the United States Government nor any agency thereof, nor any of their employees, makes any warranty, express or implied, or assumes any legal liability or responsibility for the accuracy, completeness, or usefulness of any information, apparatus, product, or process disclosed, or represents that its use would not infringe privately owned rights. Reference herein to any specific commercial product, process, or service by trade name, trademark, manufacturer, or otherwise does not necessarily constitute or imply its endorsement, recommendations, or favoring by the United States Government or any agency thereof. The views and opinions of authors expressed herein do not necessarily state or reflect those of the United States Government or any agency thereof.

## CONTACT INFORMATION

Your comments and questions are valued and encouraged. Contact the author at:

Rupom Bhattacherjee
Oklahoma State University
+1 (405)-385-2394
rupom.bhattacherjee@okstate.edu

Kodjo Opoku Botchway
Oklahoma State University
+1 (806)-224-8004
kodjo.botchway@okstate.edu